\documentclass[]{aa} % for a referee version
\begin{document}
\outer\def\gtae {$\buildrel {\lower3pt\hbox{$>$}} \over 
{\lower2pt\hbox{$\sim$}} $}
\outer\def\ltae {$\buildrel {\lower3pt\hbox{$<$}} \over 
{\lower2pt\hbox{$\sim$}} $}
\newcommand{\ergscm} {erg s$^{-1}$ cm$^{-2}$}
\newcommand{\ergss} {erg~s$^{-1}$}
\newcommand{\ergsd} {erg~s$^{-1}$ $d^{2}_{100}$}
\newcommand{\pcmsq} {cm$^{-2}$}
\newcommand{\ros} {\sl ROSAT}
\newcommand{\exo} {\sl EXOSAT}
\newcommand{\xmm} {\sl XMM-Newton}
\newcommand{\chan} {\sl Chandra}
\def\rchi{{${\chi}_{\nu}^{2}$}}
\def\uchi{{${\chi}^{2}$}}
\newcommand{\Msun} {$M_{\odot}$}
\newcommand{\Mwd} {$M_{wd}$}
\def\Mdot{\hbox{$\dot M$}}
\def\mdot{\hbox{$\dot m$}}
\input psfig.sty

\title{{\sl Chandra} observations of the globular cluster M54}
\titlerunning{{\sl Chandra} observations of the globular cluster M54}
\authorrunning{Ramsay \& Wu}

\author{Gavin Ramsay\inst{1} and Kinwah Wu\inst{1}}

\offprints{G. Ramsay}

\institute{
$^{1}$Mullard Space Science Laboratory, University College London,
Holmbury St. Mary, Dorking, Surrey, RH5 6NT, UK}

\date{Accepted A\&A}

\abstract{We have carried out a {\chan} observation of the globular
cluster M54.  We detected 7 sources located within the half-mass
radius of M54, at a flux limit of $1.5 \times 10^{-15}$\ergscm\ in the
0.3$-$8~keV energy band. The spatial distribution and the
colour/spectral properties of the 7 sources suggest that they are
likely to be cataclysmic variables or LMXBs in the globular
cluster. M54 shows the largest number of X-ray sources with
luminosities greater than $10^{32}$\ergss compared to other globular
clusters observed using {\chan} and {\xmm}. We searched for a
correlation between the number of sources above this luminosity level
with globular cluster parameters. We found evidence that the number of
sources peaks at a King concentration parameter c$\sim$1.7-1.9, with
globular clusters which are core-collapsed or have low-c values having
a smaller number of sources. We speculate on possible reasons for
this.  }

\maketitle

\section{Introduction}

M54 is a globular cluster located in the central region of the
Sagittarius (Sgr) dwarf spheriodal galaxy (Ibata, Gilmore \& Irwin
1994), which is at a distance of 27.4~kpc (Layden \& Sarajedini 2000)
and is the second closest satellite to the Milky Way.  The host galaxy
extends over at least $22^{\circ} \times 8^{\circ}$ in the sky, and is
located on the far side of the Milky Way from the Sun (Ibata, Gilmore
\& Irwin 1994, 1995).  Some theories suggest that M54 is the original
nucleus of Sgr, but an alternative view is that the globular was
actually captured (Monaco et al.\ 2003).  In addition to M54, several
other globular clusters, Arp 2, Ter 7 and Ter 8, are also associated
with the galaxy (Ibata et al.\ 1995).  Sgr is now being disrupted by
the Milky Way, and these globulars might be the last remaining
globular members that are still `attached' to Sgr.

Observations show that there is an excess of X-ray sources in Galactic
globulars. It has been argued that processes in dense stellar
environments (e.g.\ tidal capture and three-body interaction) are
crucial in the formation and evolution of globular X-ray sources
(eg Verbunt \& Hut 1987; Johnston \& Verbunt 1996; Pooley et al.\
2003).

In 2003 we carried out a {\sl Chandra} observation of the Sgr galaxy
and the globular cluster M54 to search for X-ray sources.  Here we
report our findings and present a brief comparison of the number of
sources in M54 with those in other Galactic globulars observed using
{\sl Chandra}.
   
\section{Observations and Data Analysis}

M54 was observed using {\sl Chandra} and the ACIS-I for 30~ksec
on 2003 Sept 3.  The pointing was centered on $\alpha_{2000}=18^{h}
55^{m} 3.0^{s}$, $\delta_{2000}=-30^{\circ} 28{'} 59{''}$: this was
offset from the center of the globular cluster M54 by $-12{''}$ in
declination.  The total field of view was 16.9${'}\times16.9{'}$.
Front illuminated CCDs 0--3 were used.  The read mode was configured
in {\tt TIMING} mode; the data-mode was {\tt VERY FAINT}.  During the
observation the solar particle background was very low.

We used the primary data products as supplied by the {\chan} Data
Archive Operations. A source search was performed in the 0.3$-$8.0~keV
energy range using {\tt wavedetect} (part of the CIAO v3.0 suite of
software). This was done in conjunction with an exposure map using a
mean photon energy of 1.5~keV. We rejected all `sources' with
`significance' (as defined in the {\tt wavedetect} routine) levels
smaller than 3. The resulting source list was then compared by-eye
with the image in that band: `sources' which did not appear point-like
were removed from the source list.

\begin{figure}
\begin{center}
\setlength{\unitlength}{1cm}
\begin{picture}(6,7.3)
\put(-1.5,-2.7){\includegraphics{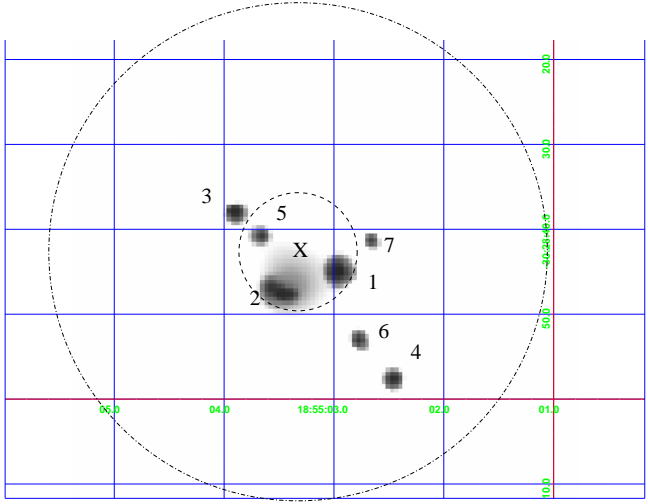}}
\end{picture}
\end{center}
\caption{ The {\chan} ACIS-I image showing the 7 sources in the
globular cluster M54. The image has been smoothed using the CIAO task
csmooth. The cluster center (derived from contour plots of the Digital
Sky Survey image of M54) is marked by an 'X'.  The core radius
(6.6$^{''}$) and the half-mass radius (29.4$^{''}$) are denoted by
concentric circles (values taken from Harris 1996).}
\label{chandra_center}
\end{figure}

Seven sources were detected in the proximity of M54 (Figure
\ref{chandra_center}) -- all of them are located well within the
cluster half-mass radius of $29.4{''}$ (Harris 1996). Source 2 could
be a blend of two sources. Another 80 sources were detected elsewhere
in the field - the `field' sources (Ramsay \& Wu, in prep).

The positions and the photon counts of the M54 sources are shown in
Table \ref{sources}.  Based on the spectral analysis described in \S
\ref{luminosity}, we estimate that the flux sensitivity limits in the
0.3$-$8~keV band is $1.5\times 10^{-15}$\ergscm\ (observed),
$2.0\times 10^{-15}$ \ergscm\ (unabsorbed) under the assumption that
the sources have a power-law spectrum of photon index $\Gamma = 1.7$
and a line-of-sight absorption column density of $1.2 \times
10^{21}$~cm$^{-2}$.  If we adopt a distance of 27.4~kpc (Layden \&
Sarajedini 2000), then the lower limit for the unabsorbed
bolometric X-ray luminosity is $1.5 \times 10^{32}$erg~s$^{-1}$.

We determined the count rates for the sources in 3 energy bands,
0.3$-$1~keV (`soft' $\equiv$ S), 1$-$2~keV (`medium' $\equiv$ M) and
2$-$8~keV (`hard' $\equiv$ H), following the convention used by
Prestwich et al.\ (2003), Soria \& Wu (2003) and Swartz et al.\ (2004)
for the analysis of properties of X-ray sources in nearby spiral
galaxies.  The counts were determined using {\tt wavedetect} and
appropriate exposure maps.  The count rates for the sources in the
three bands are shown in Table \ref{sources}. These 3 bands are used
to derive the soft and hard X-ray colours.
    
\begin{table*}
\begin{center}
\begin{tabular}{rccrrrrrr}
\hline
Id & RA & Dec & Cts 
   & Sig & Cts  & Cts & Cts & $L_{\rm x}$ 
  (10$^{33}$~erg~s$^{-1})^{*}$ \\
   &    \multicolumn{2}{c}{(2000.0)}   & (total)  &  &
(soft) & (medium) & (hard) &  (0.5$-$2.5~keV)\\
\hline
   1 &  18  55  02.96 & $-$30  28  45.0 & 93.1 & 33.3 & 36.0 & 58.0 &
5.6& 1.2\\
   2 &  18  55  03.48 & $-$30  28  47.4 & 57.0 & 17.5 & 12.4 & 25.7 &
- & 0.72\\
   3 &  18  55  03.89 & $-$30  28  38.0 & 24.1 & 10.0 &  9.7 & 14.5 &
- & 0.30\\
   4 &  18  55  02.45 & $-$30  28  57.5 & 18.0 &  7.8 &   -  &  7.8 &
9.6 & 0.23\\
   5 &  18  55  03.65 & $-$30  28  41.1 & 12.3 &  6.1 &   -  &   -  &
14.2 & 0.16\\
   6 &  18  55  02.76 & $-$30  28  53.0 & 10.6 &  5.5 &  3.7  &  8.5 &
- & 0.13 \\
   7 &  18  55  02.66 & $-$30  28  41.5 &  8.1 &  3.6 &   -  &   -  &
7.6 & 0.10\\
\hline
\end{tabular}
\end{center}
\caption{The sources associated with the
globular cluster M54. The significance levels were
derived using {\tt wavedetect} in CIAO.  The total energy band used
was 0.3$-$8~keV.  Soft counts refer to the 0.3$-$1~keV band (S),
medium counts the 1$-$2~keV band (M) and hard counts the 2$-$8~keV
band (H), following the convention of Prestwich et al.\ (2003).
$^{*}$ In determining the X-ray luminosities we adopted a distance
of 27.4~kpc and made a count to flux conversion based on spectral
fitting (see the description in \S 4).}
\label{sources}
\end{table*}

For an assumed distance of 27.4~kpc the brightest source in the M54
field has an X-ray unabsorbed luminosity of $\sim 2 \times
10^{33}$\ergss (in the 0.1--10keV energy band). This value is similar
to those sources detected in the globular clusters 47~Tuc, NGC~6397,
NGC~6440 and NGC~6752. Using the method of Wheatland (2004) we
determined the slope of the source luminosity function to be
$0.90\pm0.34$ using a lower cut-off of 8 counts. The slope of the
luminosity function of the M54 sources is steeper than those of the 4
globulars mentioned above (cf Pooley et al.\ 2002b).

If the 80 field sources are a mixture of foreground and
background sources, then we expect 0.2 field sources within the
cluster half mass radius, or 0.01 within the core radius. This is
similar to the predicted number of background AGN sources: the results
of Rosati et al (2002) predict 0.3 and 0.01 AGN within the cluster
half mass and core radius respectively in the 2--10keV energy
band. Further, if we remove all the sources detected in the whole
field and sum up the remaining number of photons, we find that we
expect 3.9 photons within an area equivalent to the cluster core
radius or 0.029 photons per square arcsec.  We conclude that the 7
X-ray sources are physically associated with the globular cluster.

\section{X-ray colours}
\label{spec}  

\begin{figure}
\begin{center}
\setlength{\unitlength}{1cm}
\begin{picture}(6,6.5)
\put(-1.5,-0.2){\includegraphics{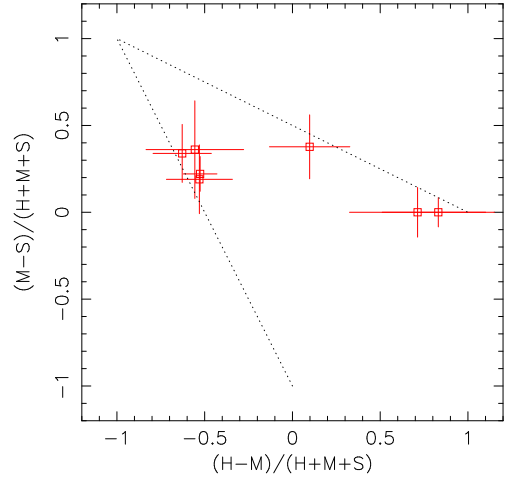}} 
\end{picture}
\end{center}
\caption{ The X-ray colour-colour diagrams of the sources in the
globular cluster M54.  The error bars represent the uncertainties due
to Poisson statistics of photon counts in the S (0.3$-$1~keV), M
(1$-$2~keV) and H (2$-$8~keV) bands.  The dotted lines enclose the
possible colour-colour tracks for the models outlined in the
text. (The abrupt change in the colour-colour track at -1,1 is
due to the relatively coarse sampling of the line of sight
absorption).}
\label{colour_colour}
\end{figure}

We derived the soft and hard colours ($\xi \equiv$\ (M$-$S)/(H+M+S)
and $\eta \equiv$\ (H$-$M)/(H+M+S) respectively) for the sources
following the convention used by Prestwich et al.\ (2003), Soria \& Wu
(2003) and Swartz et al.\ (2004). Figure \ref{colour_colour} shows the
M54 sources in the colour-colour plane.
    
\begin{figure}
\begin{center}
\setlength{\unitlength}{1cm}
\begin{picture}(6,6.5)
\put(-1.5,-0.2){\includegraphics{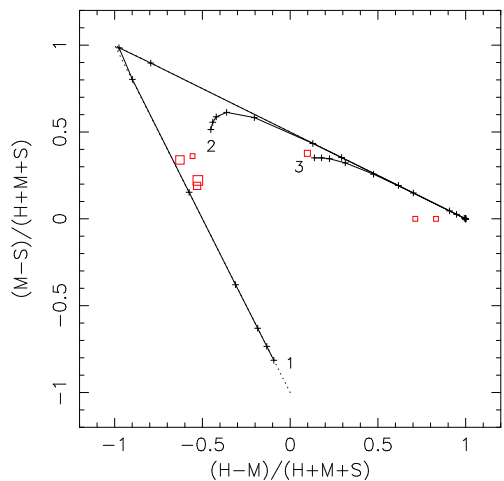}} 
\end{picture}
\end{center}
\caption{ The colours of the M54 sources in comparison with tracks for
the blackbody model with different temperatures and absorption column
densities.  The size of the symbol is proportional to the number of
counts in the 0.3$-$8~keV band.  Tracks 1, 2 and 3 correspond to the
temperatures $kT_{\rm bb} = 0.1$, 0.5 and 1.0~keV respectively.  The
absorption column density $n_{\rm H}$ starts at 1.2$\times10^{21}{\rm
cm}^{-2}$ (the column density to the edge of the Galaxy in the Sgr
direction), then steps through 5$\times10^{21}$, 1$\times10^{22}$
\pcmsq etc. For very high absorption column densities, the three
colour tracks converge to ($1,0$) in the colour-colour plane.}
\label{colour_bb}
\end{figure} 

We compared the colours of the sources with 3 spectral models: a
blackbody, a power-law and an optically thin thermal plasma.  In each
model, we considered absorption column densities ranging from the
Galactic hydrogen absorption density to the edge of the Galaxy ($1.2
\times 10^{21}{\rm cm}^{-2}$) to a column density of $N_{\rm H} =
10^{24}$~cm$^{-2}$.  For the blackbody model we considered a
temperature $kT_{\rm bb}=$100~eV, appropriate for super-soft sources,
and $kT_{\rm bb} = 0.5$ and 1~keV, similar to the inner
accretion-temperatures of X-ray binaries. For the power-law model we
considered photon indices $\Gamma=1.4$ and 1.7, typical for X-ray
binaries in the hard X-ray spectral state.  The optically thin thermal
plasma model we considered ($kT_{\rm th} = 10$~keV), is the
temperature typical of the accretion shocks in magnetic CVs (mCVs).
We also considered two lower temperatures (0.5 and 1.0~keV) for hot
plasmas in supernova remnants and coronal stars. As an example, we
show the colour-colour tracks of the blackbody models in Figure
\ref{colour_bb}. The other models follow similar colour tracks.

All the X-rays sources show colours with $-0.2 <$(M$-$S)/(H+M+S).
Soria \& Wu (2003) show that this region in the colour-colour diagram
is populated by a collection of objects which may have a
blackbody-like spectrum of $kT_{\rm in} \sim 1$~keV; an absorbed
power-law spectrum with $\Gamma \sim 1.4 -1.7$; or a spectrum of a hot
($kT_{\rm th} \sim 10$~keV) optically thin thermal plasma.  Objects
such as accreting X-ray binaries and cataclysmic variables show such
spectra.

\section{Luminosity estimates}
\label{luminosity}      

Estimating the luminosity of faint X-ray sources is not
straight-forward. However, we extracted a single spectrum from the 3
brightest soft M54 sources and a single spectrum from a field source
which was hard and relatively bright. We then fitted these spectra
using a soft blackbody model and a blackbody plus power law model
respectively (in addition to a neutral absorption model with
absorption $1.2\times10^{21}$ \pcmsq). We then determined a mean
counts to flux conversion; 1 ct gives an unabsorbed flux of
1.4$\times10^{-16}$ \ergscm \hspace{1mm} in the 0.5--2.5keV energy
band. Applying this mean conversion, instead of using either
the soft or hard spectral model, gives an uncertainty of 30\% in the
luminosity. Extrapolating the spectra to an energy band of
0.1--10keV increases the inferred unabsorbed luminosity by a factor
of 1.3 (for soft sources) and 4.5 (for hard sources).

Based on these conversions, we estimated the luminosity of the M54
sources assuming a distance of 27.4~kpc.  The sources have
luminosities in the range $\sim 0.1 - 1.2 \times10^{33}$\ergss\ in the
0.5$-$2.5~keV band. The luminosities of the sources are tabulated in
Table \ref{sources}.

The luminosity estimates suggest that the M54 sources are unlikely to
be main sequence stars, young stellar objects, RS CVn or Algol stars,
which generally have X-ray luminosities less than $10^{32}$\ergss\
(see Table 1 of Muno et al.\ (2003) and references therein).  The
source luminosities are consistent with those of mCVs, which have
luminosities $\sim 10^{31-33}$\ergss.  They are also in the range
allowed for low mass X-ray binaries (LMXBs) and high mass X-ray
binaries (HMXBs), which contain an accreting neutron star or black
hole which are in quiescent accretion states. They are unlikely to be
HMXBs since the stars in globular clusters are old. We suggest that
the M54 sources are most probably CVs or LMXBs.

\section{Comparison with Other Globular Cluster Sources} 

A number of globular clusters have been observed using {\chan} and
{\xmm} (e.g. M80, see Heinke et al.\ 2003b).  These observations show
that the number of X-ray sources scales with the encounter rate
(Pooley et al.\ 2003).  Moreover, the source luminosities suggest that
most of these systems are CVs, which are formed via a close encounter
between a white dwarf and a low-mass star or a binary (see e.g.\
Verbunt 2004).

We have identified 7 sources within the half-mass radius of M54.  All
7 sources have inferred luminosities greater than $\sim10^{32}$\ergss\
in the 0.5$-$2.5~keV band. We compare the number of sources in M54
with those of other globular clusters in the same energy
band. For those systems where it was possible to make this comparison
we show in Table \ref{compx} the number of sources in each cluster
which have luminosities greater than $10^{32}$\ergss\hspace{1mm}
(0.5-2.5keV band). Compared to the globular clusters noted in Table
\ref{compx}, M54 has more sources with luminosities above
1$\times10^{32}$\ergss.

\begin{figure}
\begin{center}
\setlength{\unitlength}{1cm}
\begin{picture}(6,10)
\put(-1.5,-0.8){\includegraphics{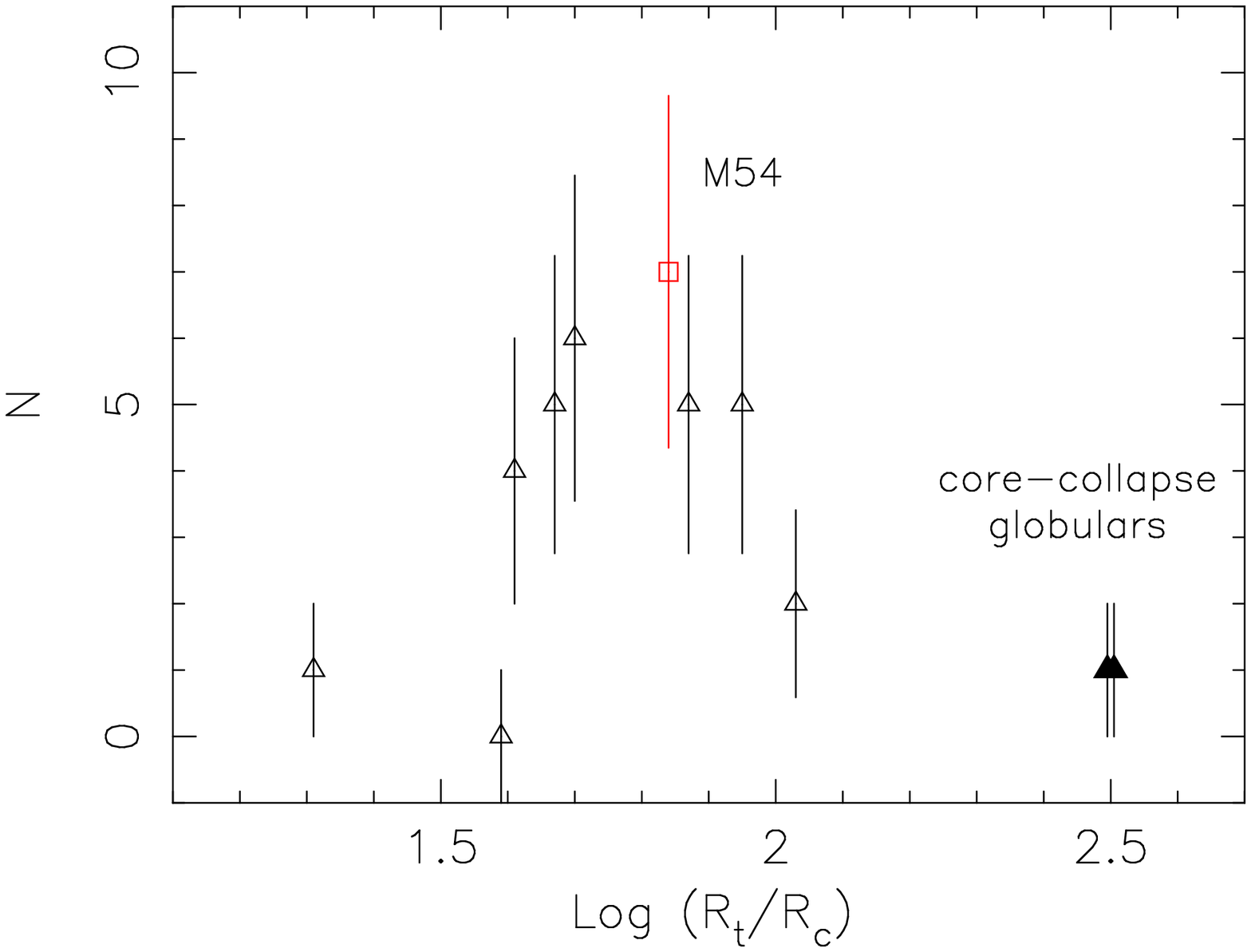}} 
\put(-1.5,4.5){\includegraphics{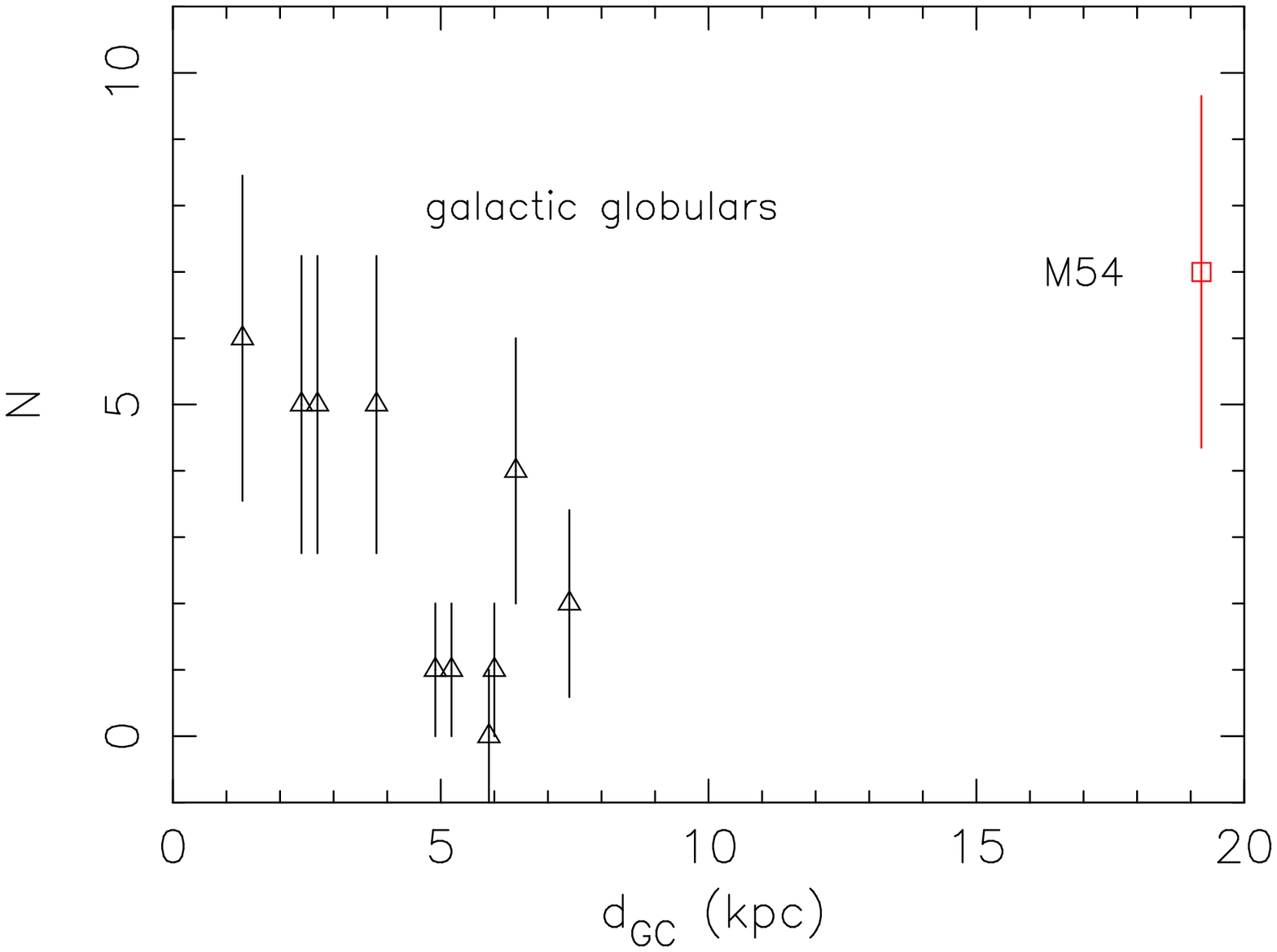}} 
\end{picture}
\end{center}
\caption{ (Top panel): The number of sources in globular clusters with
$L_{X}>1\times10^{32}$ \ergss (0.5--2.5keV) as a function of distance
of the cluster to the Galactic center.  (Bottom panel): The number of
sources in globular clusters as a function of the King (1966)
concentration.  The core-collapsed globular clusters, which are
represented by filled triangles, are assumed to have $c \equiv \log
[r_{\rm t}/r_{\rm c}] = 2.5$.}
\label{gc_1}
\end{figure}

M54 is one of the most optically luminous globular clusters associated
with the Milky Way or its satellites (Harris 1996).  We searched for a
correlation between the number of X-ray sources above
1$\times10^{32}$\ergss\ and the system parameters of the host globular
cluster.  We considered only global parameters measurable by
observations or derived directly and explicitly from observations.  We
ignored local parameters or derived parameters which are based on
models which invoke implicit assumptions.  We found no convincing
evidence for a correlation between the source number and the central
surface brightness, the core radius, the half-mass radius, the tidal
radius, the relaxation time, the metallicity or the distance to the
Galactic center (see Fig.\ \ref{gc_1}).

However, when we plotted the source number against the King (1966)
concentration ($c = \log [r_{\rm t}/r_{\rm c}]$, where $r_{\rm t}$ is
the tidal radius and $r_{\rm c}$ is the core radius), we noticed that
the source number peaks at $c \approx 1.7 - 1.9$.  Core-collapsed
globular clusters and low-$c$ globular clusters appear to have fewer
sources brighter than $\sim 10^{32}$\ergss\ (Fig.\ \ref{gc_1}).  A
simple KS test (against a uniform distribution with respect to $c$)
yielded a $D$-statistic of 0.356.  The critical $D$ value at the
significant level $\alpha = 0.01$ is 0.268.  Thus, the peak at $c
\approx 1.8$ is unlikely to be due to statistical fluctuations.

If we accept that the peak is genuine, this may indicate that too high
a concentration of stars in the core of a globular cluster reduces the
number of binary X-ray sources. This could be due to the
formation rate of binaries being less than the dissociation rate;
another possibility could be that a substantial fraction of binaries
have been ejected from the high-density core of the collapsed globular
clusters (R. Taam, private communication). More globular clusters
(especially those with large and small $c$) are needed to fully
confirm this result.

\begin{table}
\begin{center}
\begin{tabular}{lrr}
\hline
Cluster & Sources & Reference\\
\hline
NGC 6440 & 6 & (1)\\
Terzan 5 & 5 & (2)\\
47 Tuc & 2 & (3)\\
M80 & 5 & (4)\\
M28 & 5 & (5)\\
NGC 6752 & 1 & (6)\\
M4 & 0 & (7)\\
NGC 6397 & 1 & (8)\\
M22 & 1 & (9)\\
$\omega$ Cen & 4 & (10) \\
\hline
\end{tabular}
\end{center}
\caption{
The number of X-ray sources in different globular clusters
   which have luminosities greater than $1\times10^{32}$\ergss
   (observed) or $2.5\times10^{32}$\ergss (unabsorbed),  
   depending on what is quoted in the literature. 
References: (1) Pooley et al.\ (2002b); 
                       (2) Heinke et al.\ (2003a); 
                       (3) Grindlay et al.\ (2001a); 
                       (4) Heinke et al.\ (2003b); 
                       (5) Becker et al.\ (2003); 
                       (6) Pooley et al.\ (2002a); 
                       (7) Bassa et al.\ (2004); 
                       (8) Grindlay et al.\ (2001b); 
                       (9) Webb, Gendre \& Barret (2002); 
                       (10) Gendre, Barret \& Webb (2003).
} 
\label{compx}
\end{table} 

\section{Summary}

We have undertaken an X-ray survey of the globular cluster M54.  We
search down to a flux limit of $1.5 \times 10^{-15}$\ergscm in the
$0.3-8$~keV band.  A total of 7 sources were identified as located
within half-mass radius of M54. The over-density of sources within the
half-mass radius of M54 in comparison with the field sources indicates
that the 7 sources belong to the globular cluster and are not likely
to be foreground stars or background AGN.
   
The colours and the inferred luminosities suggest that the M54 sources
are accreting systems, likely to be magnetic CVs or LMXBs. By
comparison with other globular clusters observed by {\chan} and
{\xmm}, M54 is among those with most `luminous' X-ray sources (with
$L_{\rm x} > 10^{32}$\ergss).  There appears to be a range of central
concentration ($c \approx 1.7-1.9$) where the globular clusters have a
significant number of bright sources.

\section{Acknowledgments}

We thank Michael Wheatland for correspondence and discussion regarding
Bayesian statistics. We thank Allyn Tennant for correspondence in
source analysis and Ron Taam for discussion in X-ray sources in
globular clusters. We also thank Doug Swartz and Kajal Ghosh for
discussions and the referee for useful comments.

\end{document}